\documentclass[a4paper]{jpconf}

\usepackage{graphicx}

\def\lsim{\raise0.3ex\hbox{$\;<$\kern-0.75em\raise-1.1ex
\hbox{$\sim\;$}}}
\def\gsim{\raise0.3ex\hbox{$\;>$\kern-0.75em\raise-1.1ex
\hbox{$\sim\;$}}}

\begin{document}
\title{Large Extra Dimensions and Neutrino Oscillations}
\author{P.~A.~N.~Machado$^{1,2}$, H.~Nunokawa$^3$, F.~A.~ Pereira dos
  Santos$^3$, R.~Zukanovich Funchal$^1$} 

\address{$^1$Instituto de F\'{\i}sica, Universidade de S\~ao Paulo, 
  C.\ P.\ 66.318, 05315-970 S\~ao Paulo, Brazil  }
\address{$^2$ Institut de Physique Th\'eorique,
  CEA-Saclay, 91191 Gif-sur-Yvette, France  }
\address{$^3$ Departamento de F\'{\i}sica,
  Pontif{\'\i}cia Universidade Cat{\'o}lica do Rio de Janeiro,
  C. P. 38071, 22452-970, Rio de Janeiro, Brazil  }

\ead{accioly@fma.if.usp.br}

\begin{abstract}
We consider a model where right-handed neutrinos propagate in a large
compactified extra dimension, engendering Kaluza-Klein (KK) modes,
while the standard model particles are restricted to the usual
4-dimensional brane. A mass term mixes the KK modes with the standard
left-handed neutrinos, opening the possibility of change the 3
generation mixing pattern. We derive bounds on the maximum size of the
extra dimension from neutrino oscillation experiments. We show that
this model provides a possible explanation for the deficit of $\nu_e$
in Ga solar neutrino calibration experiments and of the $\bar\nu_e$ in
short baseline reactor experiments.\\
\noindent\emph{Contribution to NUFACT 11, XIIIth International
    Workshop on Neutrino Factories, Super beams and Beta beams, 1-6
    August 2011, CERN and University of Geneva}\\ 
(Submitted to IOP conference series)\end{abstract}

\section{Introduction}
\label{sec:intro}
Our purpose is to study a large extra dimension (LED)
scenario~\cite{ADD} where right (left) handed neutrinos
(antineutrinos), i.e., standard model singlet fermions, can propagate
in the 1+3+$\delta$-dimensional spacetime, while all the other
Standard Model (SM) fields are limited to the 1+3 dimensional
brane. As discussed in
Refs.~\cite{Dienes:1998sb,ArkaniHamed:1998vp,Dvali:1999cn,
  Barbieri-et-al,Mohapatra-et-al}, the neutrino masses arise from the
Yukawa coupling between the standard left-handed neutrino and the
singlet fermions in the 1+3 dimensional brane. Although this Yukawa
coupling naturally explains the smallness of neutrino masses via a
volume suppression, it also induces mixing among the standard
neutrinos and the KK modes that arise from the singlet fermions after
dimensional reduction \cite{mnz2010}. Hence, the effect of LED in
neutrino oscillations could be probed by terrestrial neutrino
experiments. There are plenty data from solar, atmospheric and
terrestrial neutrino experiments which can be very well described with
the standard three flavor oscillation scheme (see
\cite{Nakamura:2010zzi} and references therein). Being so, LED, if
present, is expected to contribute at the most as a subdominant effect
on top of the usual oscillation pattern.  Nevertheless, a few mild
deviations from that scheme remain.

First, the LSND experiment~\cite{LSND} has observed an unexpected
excess of $\bar{\nu}_e$ events in the $\bar{\nu}_\mu \to \bar{\nu}_e$
mode, which is also supported by more recent MiniBOONE
data~\cite{MiniBOONE}. Moreover, the GALLEX~\cite{gallex} and
SAGE~\cite{sage} experiments, designed to calibrate gallium
radiochemical solar neutrino detectors, observed a smaller than
expected $\nu_e$ flux originated from $^{51}$Cr (GALLEX and SAGE), and
$^{37}$Ar (SAGE only). The mean value of the ratios of the measured
over predicted rates is $0.86\pm 0.05$, comprising a
$2.7\sigma$ deviation~\cite{giunti2}, which is the so-called {\em
  gallium anomaly}. Finally, a recent re-evaluation of the reactor
$\bar \nu_e$ flux~\cite{Mueller:2011nm,Huber:2011wv} resulted in an
increase in the flux of 3.5\%. Although the impact on the results of
long baseline experiments such as KamLAND is negligible, the new flux
calculations induce an average deficit of 5.7\% in the observed event
rates for short baseline ($ < 100$ m) reactor neutrino experiments.
This constitute a 98.6 \% CL deviation from unity and has been
referred to as the {\em reactor antineutrino
  anomaly}~\cite{reactor-anomaly}.

In this work, we will derive bounds on the size of the largest extra
dimension from neutrino oscillation experiments and show that the
discussed anomalies can spring from a LED
model.

\section{Neutrino Oscillations with a Large Extra Dimension}
\label{sec:LED}

The LED model considered here is the one described in
Refs.~\cite{mnz2010,Davoudiasl:2002fq}. In this scheme, the SM fields
are confined to the 1+3-dimensional brane and three SM singlet fermion
field can propagate in the higher dimensional bulk with at least two
compactified flat extra dimensions.  Let us say that one of these
extra dimensions is compactified on a circle of radius $a$ and, for
simplicity, is much larger than the others, so that a 5-dimensional
treatment is a good approximation.  The 3 bulk fermions have Yukawa
couplings with the SM Higgs and the brane neutrinos ultimately leading
to flavor oscillations driven by Dirac masses, $m_i$ ($i$=1,2 and 3),
and KK masses $m^{\rm KK}_n$ ($n=1,2,...$), and mixings among active
species and sterile modes. In this case the survival probability
$\nu_\alpha\to\nu_\alpha$ ($\alpha=e,\mu,\tau$) in vacuum, which holds
also for antineutrinos, can be approximated
by~\cite{mnz2010,Davoudiasl:2002fq,mnpz2010}

\vglue -0.6cm
\begin{equation}
P(\nu_\alpha \to \nu_\alpha;L,E) = \vert {\cal{A}}_{\nu_\alpha\to
  \nu_\alpha} (L,E)\vert^2 \, , \quad\quad {\rm with} \quad\quad
{\cal{A}}_{\nu_\alpha \to \nu_\alpha} (L,E) = \sum_{i=1}^{3} \vert
U_{\alpha i}\vert^2 A_i,
\end{equation}
\vglue -0.2cm
where $A_i$ is given by, assuming $m_ia \ll 1$ and ignoring the terms
of order $(m_ia)^3$ and higher in the amplitude as well as $(m_ia)^2$
and higher in the phase,
\begin{equation}
\hskip -1cm A_i \approx (1 - \frac{\pi^2}{6} m_i^2 a^2)^2 \exp \left(i
\frac{m_i^2 L}{2E}\right) + \sum_{n=1}^{\infty} 2
\left(\frac{m_i}{{m_n^{\rm KK}}}\right)^2 \exp \left[i
  \frac{(2\,m_i^2+{m_n^{\rm KK}}^2) L}{2E}\right].
\label{eq:amplitude}
\end{equation}
Here $U_{\alpha i}$ are the elements of the usual Maki-Sakata-Nakagawa
neutrino mixing matrix (we use the standard parameterization found in
Ref.~\cite{Nakamura:2010zzi}), $E$ is the neutrino energy, $L$ is the
baseline distance, $m^{\rm KK}_n =n/a$ is the mass of the $n$-th KK
mode.

\begin{figure}[bhtp]
\begin{minipage}[b]{2in}
\begin{center}
\includegraphics[width=2in]{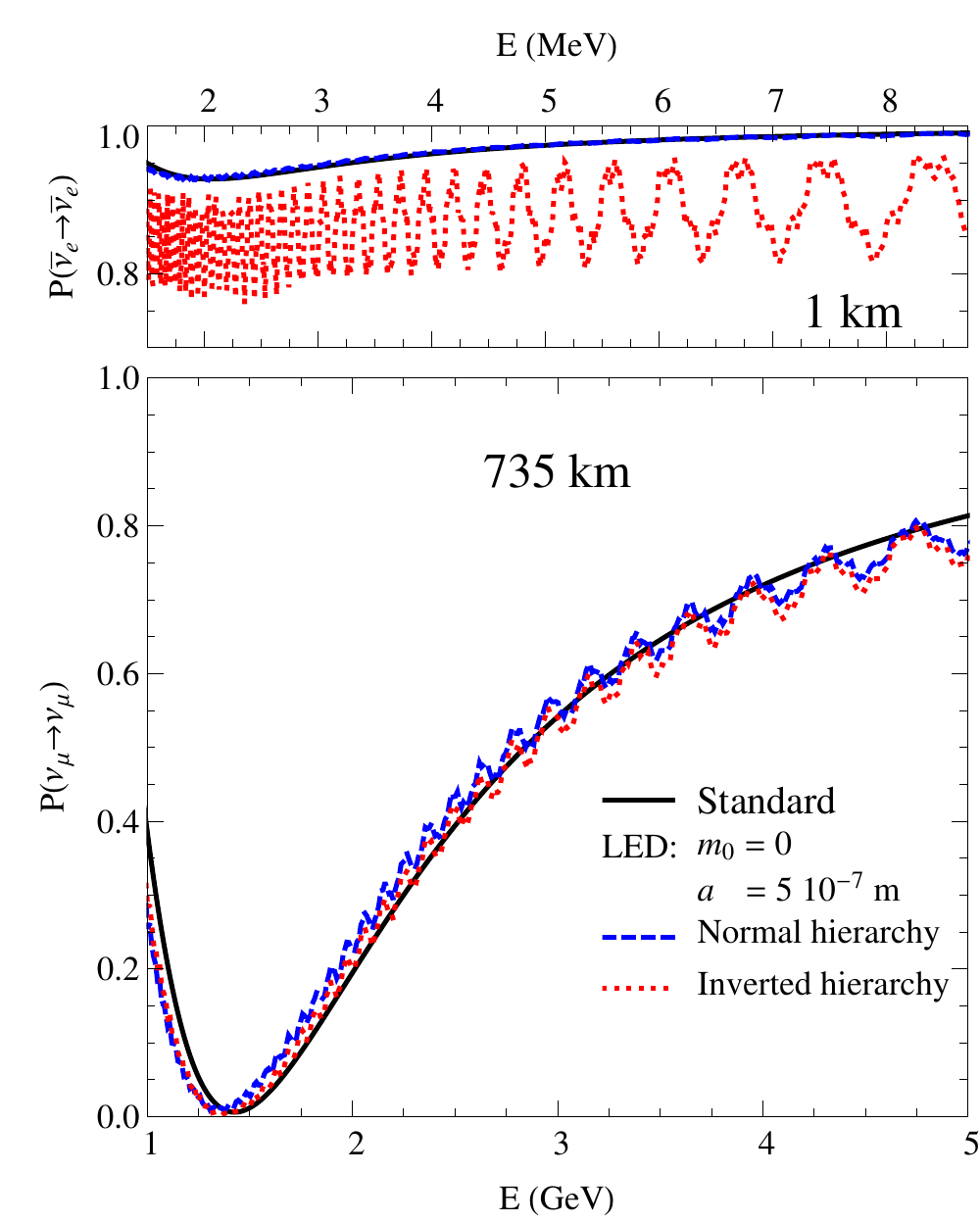}
\end{center}
\caption{Illustrative oscillation probabilities. See legend in figure
  for details.\\ \\}
\label{fig:probs}
\end{minipage}
\hspace{.2in}
\begin{minipage}[b]{4in}
\includegraphics[width=2in]{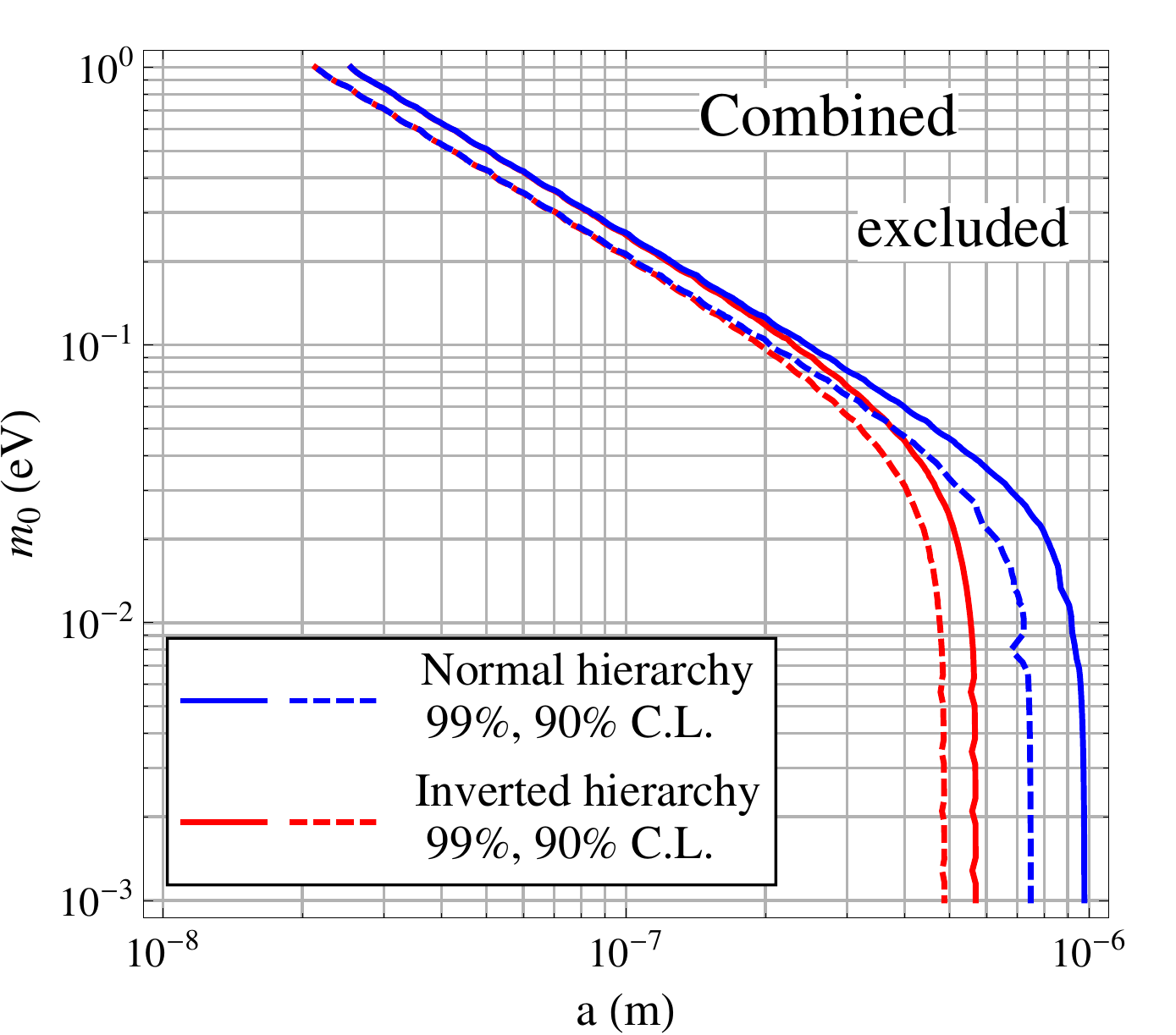}
\includegraphics[width=2in]{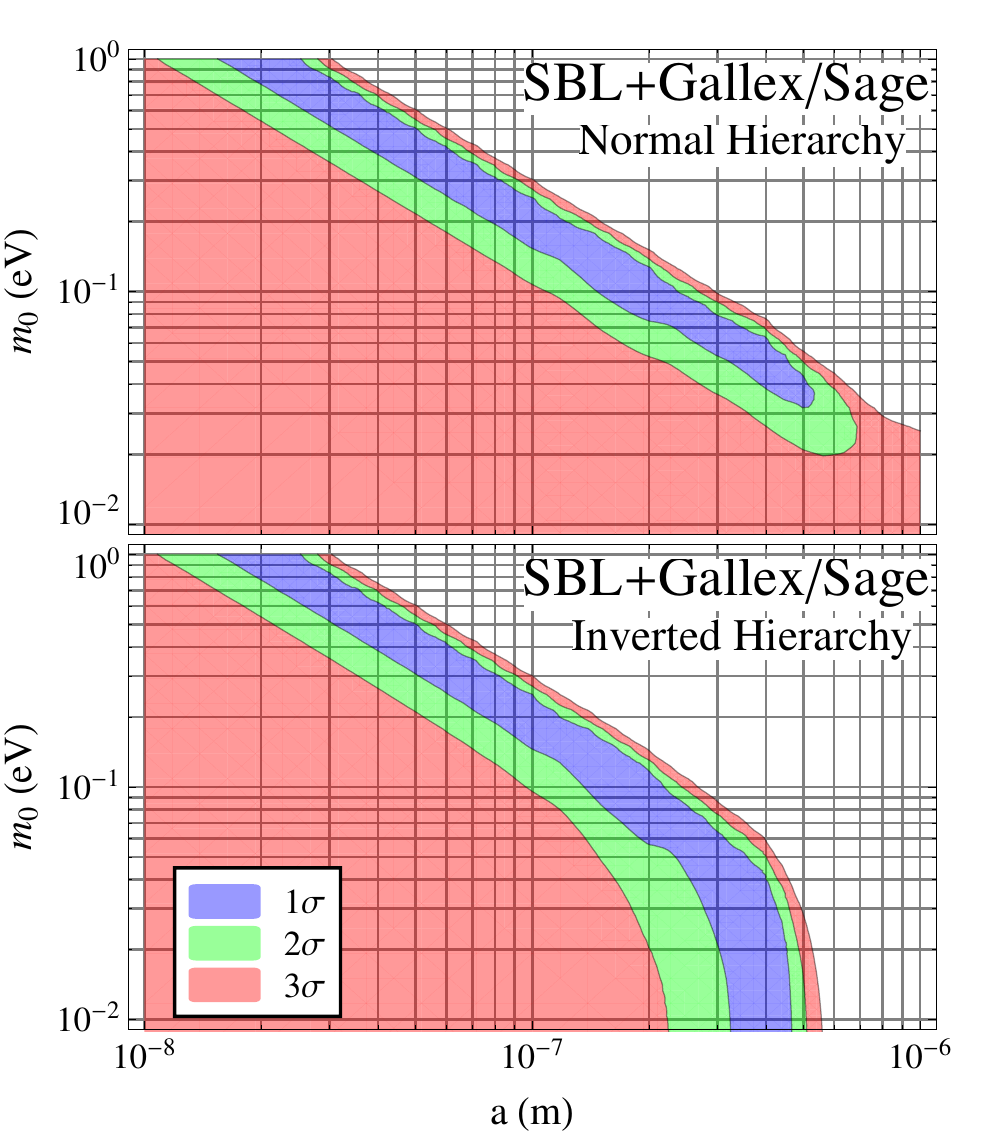}
\caption{Left: Excluded region in the $a - m_0$ plane by the combined
  CHOOZ, KamLAND, and MINOS $\nu_\mu \to \nu_\mu$ as indicated. Right:
  Allowed region in the $a - m_0$ plane by the combined data set from
  GALLEX, SAGE and short baseline reactor experiments for each
  hierarchy as indicated. }
\label{fig:comb-led}
\label{fig:allowed-regions}
\end{minipage}

\end{figure}

As can be seen, the probability depends on the absolute neutrino
masses, hence the behavior is different for normal mass hierarchy
(NH), where $m_3 > m_2 > m_1=m_0$, and inverted mass hierarchy (IH),
where $m_2> m_1 > m_3=m_0$.  Clearly, as $m_0$ increases the
differences between the hierarchies fade away and the masses become
degenerate. Thus, in this model, neutrino oscillations will be
sensitive to the standard oscillation parameters, the size of the
extra dimension, and the mass of the lightest neutrino. In
fig.~\ref{fig:probs} we show illustrative oscillation
probabilities. For a discussion about the oscillation behavior, see
Refs.~\cite{mnz2010,mnpz2010}.

We analyzed what region in the $a - m_0$ plane can be excluded by the
latest data from CHOOZ~\cite{chooz}, KamLAND~\cite{kl}, and
MINOS~\cite{minos} $\nu_\mu \to \nu_\mu$, for both hierarchies. This
was calculated at 90\% (99\%) CL, by imposing $\chi^2>\chi^2_{\rm min}
+ 4.61 \,(9.21)$. All parameters were varied freely. The combined
exclusion is shown in fig.~\ref{fig:comb-led}. See Ref.~\cite{mnz2010}
for details.

\section{An Interpretation for the Gallium and Reactor Anomalies}

Now we address the question: \emph{Can the gallium and reactor
  anomalies be due to this large extra dimensions model?} To answer
that, let us comment first on the gallium anomaly. The radiochemical
solar neutrino experiments GALLEX and SAGE have been calibrated with
monoenergetic $\nu_e$'s from intense radioactive sources, which are
captured by the reaction,
\begin{equation}
\nu_e + {\rm ^{71}Ga} \to {\rm ^{71}Ge} + e^-
\label{Eq:Ga-reaction}.
\end{equation}
As sources, the GALLEX collaboration used ${\rm ^{51}Cr}$, publishing
two measurements~\cite{gallex}, while the SAGE collaboration used both
${\rm ^{51}Cr}$ and ${\rm ^{37}Ar}$~\cite{sage}. The ratio of the
measured $^{71}{\rm Ge}$ event rate over the predicted one (using the
cross section estimated in Ref.~\cite{bahcall}) are, including
$1\sigma$ errors, for GALLEX~\cite{gallex} ($R^{\rm G}$) and
SAGE~\cite{sage} ($R^{\rm S}$).

\vglue -0.4cm
\begin{eqnarray}
R^{\rm G}_{\rm Cr1} = 0.95 \pm 0.11&,& R^{\rm S}_{\rm Cr} = 0.95 \pm
0.12,\nonumber \\ \rm R^{G}_{Cr2} = 0.81 \pm 0.11&,& R^{\rm S}_{\rm
  Ar} = 0.79 \pm 0.09.\nonumber
\end{eqnarray}
 We can see that all the ratios are below unity. Accordingly to the
 analysis in Ref.~\cite{giunti2,giunti1} this represents a $2.7\sigma$
 deviation from the expected value. We performed a similar analysis
 (see Ref.~\cite{mnpz2010} for details).
 
Now, let us discuss briefly the second anomaly, namely the reactor
antineutrino anomaly. The reactor antineutrino fluxes have been
reevaluated~\cite{Mueller:2011nm,Huber:2011wv}, which lead to a 5.7\%
decrease in the number of $\bar \nu_e$ observed and theoretically
predicted for all short baseline reactor
experiments~\cite{reactor-anomaly}.

We have simulated 19 reactor experiments with baseline shorter than
100~m~\cite{bugey3}. Our simulation follows closely the one described
in Ref.~\cite{reactor-anomaly} (for details see Ref.~\cite{mnpz2010}).
To obtain the theoretical rates with LED, we used the experimental
results available in~\cite{bugey3}
and the parameterization given in \cite{reactor-anomaly} to calculate
the expected reactor fluxes. 

We have fitted the two gallium calibration experiments described
together with the above short baseline reactor experiments using the
LED scenario discussed in Section~\ref{sec:LED}, thus obtaining the
allowed regions in the $m_0-a$ plane, for both normal and inverted
hierarchies, as shown in fig.~\ref{fig:allowed-regions}.  We found
that the combined data favor the nonzero value of the large extra
dimension, 2.9 $\sigma$ away from $a=0$. These regions quite
compatible with the limits obtained in Section~\ref{sec:LED} coming
from other oscillation experiments. The allowed region for explaining
the anomalies overlaps scarcely with those excluded by CHOOZ, KamLAND
and MINOS.

\section{Conclusions}

We investigated a large flat extra dimensions model where the SM
fields are confined to the 4-dimensional brane while three SM singlet
fermion fields can propagate in the bulk. These fields couple to the
active neutrino thru a Yukawa term. This generates small neutrino
masses, but also changes the oscillation pattern. We have shown that
terrestrial neutrino oscillation experiments can set sub-micrometer
bounds on the size of the largest extra dimension.

We also show that the gallium and reactor antineutrino anomaly can be
explained by such a model. In this case, the observed deficit of
neutrinos is due to the oscillation between active neutrinos and
sterile KK modes coming from the SM singlet fermions.  While the
future MINOS and Double CHOOZ data can improve somewhat the limits in
the small $m_0$ parameter region~\cite{Machado-NUFACT2011}, it seems
not easy to exclude or confirm the LED solution discussed in this
work.

Finally, we note that the excesses observed in the LSND and MiniBOONE
experiments can not be explained by this simple model. Although
considering two extra dimensions of different size could, in
principle, enhance short baseline $\nu_\mu \to \nu_e$ and $\bar\nu_\mu
\to \bar\nu_e$, naively there is no source of CP violation. An
explanation for LSND/MiniBOONE, gallium and reactor anomalies by some
extension of the model discussed here could be an interesting subject
for further studies.

\ack
  This work is supported by the Brazilian funding agencies
  FAPESP, FAPERJ, and CNPq and by the European Commission under
  the contract PITN-GA-2009-237920. PANM is grateful to the NUFACT
  2011 organizers for the invitation.


\section*{References}
\begin{thebibliography}{99}
\bibitem{ADD} Arkani-Hamed~N, Dimopoulos~S and Dvali~G 1998,
  \emph{Phys. Lett.} B {\bf 429} 263; 
  \emph{ibid} 1999 \emph{Phys. Rev.} D {\bf 59} 086004;
  Antoniadis~I \emph{et al}
  1998 \emph{Phys. Lett.} B {\bf 436} 257.

\bibitem{Dienes:1998sb} Dienes~K~R, Dudas~E, Gherghetta~T 1999
  \emph{Nucl.\ Phys.} B {\bf 557} 25.

\bibitem{ArkaniHamed:1998vp} Arkani-Hamed~N {\it et al.} 2002
 \emph{Phys.\ Rev.} D {\bf 65} 024032.

\bibitem{Dvali:1999cn} Dvali~G~R and Smirnov~A~Y 1999
\emph{Nucl.\ Phys.}  B {\bf 563} 63

\bibitem{Barbieri-et-al}
Barbieri~R, Creminelli~P and Strumia~A 2000
\emph{Nucl.\ Phys.} B {\bf 585} 28

\bibitem{Mohapatra-et-al}
 Mohapatra~R~N, Nandi~S and Perez-Lorenzana~A 1999
  \emph{Phys.\ Lett.}  B {\bf 466} 115;
Mohapatra~R~N and Perez-Lorenzana~A 2000
  \emph{Nucl.\ Phys.}  B {\bf 576} 466;
 {\it ibid} 2001
 \emph{Nucl.\ Phys.}  B {\bf 593} 451.

\bibitem{mnz2010} 
Machado~P~A~N, Nunokawa~H, and Zukanovich Funchal~R 2011
\emph{Phys.\ Rev.} D {\bf 84} 013003.

\bibitem{Nakamura:2010zzi}
  Nakamura~K {\it et al.}  [Particle Data Group] 2010
  \emph{J.\ Phys.} G {\bf 37} 075021.

\bibitem{LSND}
  Athanassopoulos~C {\it et al.}  [LSND Collaboration] 1996
  \emph{Phys.\ Rev.\ Lett.}  {\bf 77}, 3082;
  Aguilar~A {\it et al.}  [LSND Collaboration] 2001
  \emph{Phys.\ Rev.}  D {\bf 64} 112007

\bibitem{MiniBOONE}
  Aguilar-Arevalo~A~A {\it et al.}  [The MiniBooNE Collaboration] 2010
  \emph{Phys.\ Rev.\ Lett.}  {\bf 105} 181801

\bibitem{gallex} 
 Anselmann~P {\it et al.}  [GALLEX Collaboration] 1995
  \emph{Phys.\ Lett.}  B {\bf 342} 440;
  Hampel~W {\it et al.}  [GALLEX Collaboration] 1998
  \emph{Phys.\ Lett.}  B {\bf 420} 114;
  Kaether~F \emph{et al} 2010
  \emph{Phys.\ Lett.}  B {\bf 685} 47.

\bibitem{sage}
 Abdurashitov~J~N {\it et al.}  [SAGE Collaboration] 1999
  \emph{Phys.\ Rev.}  C {\bf 59} 2246;
 J.~N.~Abdurashitov {\it et al.} 2006
  \emph{Phys.\ Rev.}  C {\bf 73} 045805;
  J.~N.~Abdurashitov {\it et al.}  [SAGE Collaboration] 2009
  \emph{Phys.\ Rev.}  C {\bf 80} 015807.

\bibitem{giunti2} 
C. Giunti and M. Laveder 2010 \emph{Phys. Rev.} D {\bf 82} 053005;
C. Giunti and M. Laveder \emph{Preprint} 1006.3244.

\bibitem{Mueller:2011nm}
  Mueller~T~A {\it et al.} 2011
  \emph{Phys.\ Rev.}  C {\bf 83} 054615.

\bibitem{Huber:2011wv}
      Huber~P, 
      \emph{Preprint} 1106.0687.

\bibitem{reactor-anomaly} 
 Mention~G \emph{et al} 2011
  \emph{Phys.\ Rev.}  D {\bf 83} 073006.

\bibitem{Davoudiasl:2002fq} 
Davoudiasl~H, Langacker~P, and
  Perelstein~M 2002
\emph{Phys.\ Rev.}  D {\bf 65} 105015

\bibitem{mnpz2010} Machado~P~A~N \emph{et al Preprint} 1107.2400.

\bibitem{chooz}
  Apollonio~M {\it et al.}  [CHOOZ Collaboration] 2003
  \emph{Eur.\ Phys.\ J.}  C {\bf 27} 331.

\bibitem{kl} 
Gando~A {\it et al.}  [KamLAND Collaboration] 2011
 \emph{Phys.\ Rev.}  D  {\bf 83} 052002.

\bibitem{minos}  Michael~D~G {\it et al.}  [MINOS Collaboration] 2006
  \emph{Phys.\ Rev.\ Lett.}  {\bf 97} 191801; 
  Adamson~P {\it et al.}  [MINOS Collaboration] 2008
  \emph{Phys.\ Rev.\ Lett.}  {\bf 101}, 131802.

\bibitem{bahcall} 
Bahcall~J~N 1997 \emph{Phys. Rev.} C {\bf 56} 3391.

\bibitem{giunti1}  
Acero~M~A, Giunti~C, and Laveder~M 2008
  \emph{Phys.\ Rev.}  D {\bf 78} 073009


\bibitem{bugey3} 
Achkar~B {\it et al.} 1995 \emph{Nucl. Phys.} B {\bf 434} 503;
D\'eclais~Y  {\it et al.} 1994 \emph{Phys. Lett.} B {\bf 338} 383;
Kwon~H 1981 {\it et al.} \emph{Phys. Rev.} D {\bf 24} 1097;
Hoummada~A 1995 {\it et al.} \emph{Appl. Rad. Isot.} {\bf 46} 449;
Zacek~G {\it et al.}  1986 \emph{Phys. Rev.} D {\bf 34} 2621;
Greenwood~Z D {\it et al.} 1996 \emph{Phys. Rev.} D {\bf 53} 6054;
Vidyakin~G S 1987 {\it et al.} \emph{JETP} {\bf 93} 424;
Vidyakin~G S 1994 {\it et al.} \emph{JETP Lett.} {\bf 59} 390;
Afonin~A I 1988 {\it et al.} \emph{ JETP} {\bf 94} 213;
Kuvshinnikov~V 1991 {\it et al.} \emph{JETP} {\bf 54} 259.

\bibitem{Machado-NUFACT2011}
Talk given by P.A.N. Machado at 
the NUFACT 11 - CERN/UNIGE,
Geneva, Switzerland, August 1st -  6th, 2011, 
available at http://nufact11.unige.ch/

\end{thebibliography}
\end{document}